\documentstyle[12pt,aasms4]{article}
\lefthead{Gull et al.}
\righthead{NUV Spectrum of the Crab Pulsar}
\begin{document}

\title{STIS Near Ultraviolet Time-Tagged Spectra\\
    of the Crab Pulsar\altaffilmark{1}}
\altaffiltext{1}{Based on observations with the NASA/ESA \it Hubble Space
Telescope,
\rm obtained at the Space Telescope Science Institute, which is operated
by AURA Inc under NASA contract NAS5-26555.} 

\author{Theodore R. Gull\altaffilmark{2,3,4}, Don J. Lindler\altaffilmark{4,5}, D. Michael Crenshaw\altaffilmark{4,6}, Joseph F. Dolan\altaffilmark{2,4}, Stephen J. Hulbert\altaffilmark{7}, Steven B. Kraemer\altaffilmark{3,4,6}, Peter Lundqvist\altaffilmark{8}, Kailash C. Sahu\altaffilmark{7}, Jesper Sollerman\altaffilmark{8}, George Sonneborn\altaffilmark{2,4}, Bruce E. Woodgate\altaffilmark{2,3,4}}

\altaffiltext{2}{Laboratory for Astronomy and Solar Physics} 
\altaffiltext{3}{member of STIS Investigation Definition Team (IDT)}
\altaffiltext{4}{Goddard Space Flight Center, Code 681, Greenbelt, MD 20771} 
\altaffiltext{5}{Advanced Computer Concepts}
\altaffiltext{6}{Catholic University}
\altaffiltext{7}{Space Telescope Science Institute, 3700 San Martin Drive, 
Baltimore, MD 21218}
\altaffiltext{8}{Stockholm Observatory, SE-133 36, Saltsj\"obaden, Sweden}

\begin{abstract}

We present the spectrum and pulse profile of the Crab Pulsar in the near ultraviolet ($1600-3200$~\AA) observed with the Space Telescope Imaging Spectrograph (STIS) during the Hubble Space Telescope (HST) second Servicing Mission Orbital Verification (SMOV) period. The two-dimensional Near-Ultraviolet Multi-Anode Microchannel Array (NUV MAMA) was used in time-tag mode with a $2\arcsec \times 2\arcsec$ aperture and the low dispersion grating, G230L, to obtain a cube with axes of slit position, wavelength, and time. The observation-derived pulse profile is consistent with radio measurements, and the pulse profile agrees well with previous NUV broadband measurements by the High Speed Photometer. The pulsar spectrum includes the $2200$~\AA\ dust absorption feature, plus several interstellar absorption lines. Dereddening the spectrum using the Savage-Mathis model with $E(B-V)=0.55 \pm 0.05$ leads to a good fit to a power law with slope $ \alpha _{\nu}= -0.3 \pm 0.2 $. Spectra of the main pulse, the interpulse, and the individual rising and falling edges are similar to the total spectrum within the limits of photon statistics. The four pulse profiles produced by breaking the spectrum into $400$~\AA\ bins show the pulse profile to be stable across the NUV spectral range. Histogram analysis reveals no evidence for the superpulses seen at radio wavelengths. The interstellar absorption line equivalent widths of Mg I, Mg II and FeII are lower than expected based upon the implied HI column density from $E(B-V)=0.5$. While several explanations are possible, additional studies will be necessary to narrow the options.

\end{abstract}
 

\section{Introduction}

The Crab Pulsar is a recognized standard for astronomical timing measurements (\cite{lyn97}). As we prepared to demonstrate the performance of the STIS after installation in HST in 1997 February (\cite{kim98}), we found the Crab Pulsar to be uniquely suited for testing the time-tagging capability of the STIS MAMA detectors. The Crab Pulsar is well studied at multiple wavelengths, and is the only periodic rapid variable in the ultraviolet that is bright enough to record, with good signal to noise ratio, the light curve in a few HST orbits. The arrival time and period are monitored regularly at radio wavelengths (\cite{lyn97}). The Crab Pulsar pulse profile has been studied at ultraviolet and visible wavelengths (\cite{per93}) with the High Speed Photometer (HSP) on the Hubble Space Telescope, in the near infrared (\cite{eik96}, \cite{eik97}) and at radio and gamma-ray wavelengths (\cite{lun95}, \cite{tom97}). While giant pulses, as much as 60 to 2000 times the average main pulse, are detected at radio wavelengths (\cite{lun95}), Percival et al. (1993) found no evidence for giant pulses in the near ultraviolet or visible spectral regions.

Spectroscopy of the Crab Pulsar in the ultraviolet has been limited to two observations: by the International Ultraviolet Explorer (IUE) (\cite{ben80}) extending from $2000$ - $3000$~\AA\ with low signal-to-noise, and an unpublished Faint Object Spectrograph (FOS) spectrum extending from $2000$ - $6000$~\AA\ in the HST archives. Neither are sufficiently deep to provide useful information on the interstellar absorption lines and provide only limited information on the $2200$~\AA\ diffuse absorption feature due to interstellar dust.

\section{Observations}

In the ultraviolet, with the NUV MAMA and the Far Ultraviolet (FUV) MAMA, we are able to observe transient phenomena with good spatial resolution ($0\farcs025$ /pixel), good time resolution ($125~\mu$s) and the spectral resolution appropriate to the observation of interest (\cite{kim98}). In the NUV, the following modes are possible: direct imagery with a limited set of filters, long slit spectroscopy at low and moderate dispersions (R=600 and 10,000) and small slit spectroscopy at high dispersions (R=46,000 and 114,000) (\cite{hst96}). This observation is an example of the time-tagging capabilities of STIS in the ultraviolet at low resolution.

STIS was used to observe the Crab Pulsar on 1997 August 7 for three orbits using the G230L grating with the NUV MAMA detector. As we were testing the ability of the time-tag mode versus the on-board accumulate (ACCUM) mode, observations during the first orbit were recorded in ACCUM mode for three 600 s accumulations and observations during the remaining two orbits were recorded in time-tag mode. The observation log is listed in Table 1. To the limit of photon statistics, the spectra recorded in the two modes are identical. As target acquisition had not been adequately demonstrated when the observation was prepared, a conservative $2\arcsec \times 2\arcsec$ aperture was specified. By the time the observation occurred, target acquisition using the CCD camera was routine and transfer to the MAMA modes was well proven. The pulsar was indeed well centered in the slit. Wavelengths are measured to within 1/2 low resolution (lores) pixel or $0.8$~\AA. The NUV MAMA has a higher than intended dark count rate due to phosphorescence in the window; the dark count rate for the entire detector was about 1,700 counts per second. By comparison, the Crab Pulsar count rate was about 70 per second (or two per individual period). However, the pulsar spectrum is dispersed over 1024 elements with a spatial resolution determined by the telescope point spread function in the $1600$ - $3200$~\AA\ region. With the full width at half maximum of $0\farcs05$ and low scatter of the optics in this spectral region, virtually all of the photon events from the pulsar land on an eleven pixel wide spatial slice, which is $1\%$ of the total detector format. The detector dark count for the Crab Pulsar is thus about seventeen counts per second, or one fourth of the signal.

\placetable{tbl-1}

\section{Data Reduction}

As this was a first time observation with time-tag mode, all data analysis was done using IDL software tools developed for testing of the STIS before launch. We derived the pulse profile in the following manner (\cite{lin97}): a two-dimensional ACCUM image was constructed by summing up the raw time-tagged events (time, X, Y) for each X and Y coordinate of the MAMA detector. This image served as a template to select time-tag events that are associable with the dispersed pulsar spectrum. Each event arrival time was corrected to the solar-system barycenter arrival time. We then selected photon events spatially located within 5 pixels ($0\farcs125$ along the slit) of the pulsar spectrum's peak. We computed the period by maximizing the sum of the squares of the values in the pulse profile divided into 512 time bins. This resulted in a measured period at the time beginning with observation O45701C0M at time Modified Julian Day (MJD) 2450667.70478 of $33.473313$ ~ms with a statistical error being less than the last digit. The computed period based upon Jodrell Bank Crab Pulsar timing standards (\cite{lyn97}) is identical.

The time-tag observations revealed that the initial two-dimensional image had two problems: the extracted pulse profile had an extended wing for both the primary and interpulse peaks, and the detector background showed a weak pulsar period amplitude. We determined that the flight hardware was recording the X,Y coordinates, then waiting for the next event before assembling the T, X and Y event in buffer memory. The event was recorded as T(n+1),X(n),Y(n) instead of T(n),X(n),Y(n). A simple shift of the time vector by one X,Y pair rectified both the extended wing and the background periodicity. Figure 1 demonstrates the success of this procedure by plotting the STIS pulse profile on the same graph as the HSP pulse profile (\cite{per93}), originally recorded at $21.5~\mu$s resolution in the ultraviolet ($1600-3200$~\AA\ bandpass), but resampled to 512 bins or approximately $62.5~\mu$s each. We checked for period drift across the two observing periods in adjacent orbits by plotting the photon events with phase period as demonstrated by Percival et al. (1993) in their Figure 10. No drift could be detected from the beginning of the first orbit to the end of the last orbit. 

This observation utilized the MAMA detector to provide a data cube representing T(time), X(spectrum) and Y(slit), but other MAMA applications (such as the echelle modes and the direct imaging modes) apply the detector coordinates to provide alternative spectral or spatial information. For the Crab Pulsar observation, sub-cubes can be extracted with respect to time to determine spectral properties of various phases of the main pulse, the interpulse, the bridge, rising edges and falling slopes. Other sub-cubes can be extracted, using the pulsar spatial position along the slit, to study the pulse profile for smaller slices of the spectral region. One unique representation is reproduced in Figure 2 (=Plate X). The X-axis depicts the uncalibrated spectrum, while the Y-axis depicts the pulse profile. The two-dimensional plot shows the variation of the spectrum with phase of the pulse profile. No wavelength-dependent structure is noticable to the eye.

The spectra were extracted using an eleven pixel wide extraction slit. As one orbit had been expended in the ACCUM mode immediately before the time-tag orbits, we first compared the ACCUM spectrum with the time-tag spectrum. To the limit of statistical variation in number of photon events, we found no difference in the spectral properties. To obtain the best S/N possible, we combined the time-tag spectrum with the ACCUM spectrum. The extracted and summed spectrum (Figure 3a) was flux-calibrated using the on-orbit calibrations of Collins and Bohlin (1998). Absolute fluxes are accurate to within a few percent, based upon repeatability of flux measures of the several reference stars.

\section{Results}

We examined the data cube of the photon events for variations of the pulse profile as a function of wavelength and variations of the spectrum for selected intervals of the pulse profile. In terms of the wavelength, pulse profile space of Figure 2, we collapsed the data into $400$~\AA\ bins between $1600$~\AA\ and $3200$~\AA\ and derived four independent pulse profiles. For all four intervals, the ratio of the main pulse to the interpulse, or P1/P2, is stable to well within $5\%$. We looked at the spectrum of the main pulse, the interpulse, the rising edges and the falling edges, as well as the interval above $80\%$ of the main pulse. In no case do we find changes in the spectrum that exceed those expected from photon statistics, which are at the $5\%$ level. The number of events recorded in the bridge, the portion of the pulse profile between the main pulse and the interpulse, is too few to determine accurately its spectrum. The physical event creating the near ultraviolet pulsar emission is neutral in color between $1600$ and $3200$~\AA. 
	
The net emission passing through the $2 \arcsec \times 2 \arcsec$ aperture has two components: the pulsar, or stellar emission, and the nebular emission. The point-spread function of the telescope is at its best in this spectral region (\cite{sch87}), but still has extended wings. As the NUV MAMA is a two-dimensional detector that can view slits up to $26 \arcsec$ in height, we can determine the dark count rate across the detector by measuring the detected rate outside of the $2\arcsec$ slit height (80 pixels). Five regions, each eleven pixels in height, were selected for analysis: Region A, centered on the pulsar; B1, centered 30 pixels below the pulsar; B2, centered 30 pixels above the pulsar; C1, centered 60 pixels below the pulsar; C2, centered 60 pixels above the pulsar. Here, we can utilize the periodic variability of the pulsar to estimate the nebular contribution. The average dark count will be $C=(C1+C2)/2$. The nebula plus dark count plus wings of the pulsar would be $B=(B1+B2)/2$. The pulsar profile would be $D=A-B$. We can then compute the nebular contribution to be $E=B-C-$$f$D, where $f$ is fractional contribution of the stellar point spread function. We determined the value for $f$ by computing the pulse profile for E. If $f$ is too small, residual bumps would be seen in an extracted pulse profile at the positions of the main pulse and interpulse. Likewise, if $f$ is too large, residual dips would be seen. We determined the value for $f$ to be 0.016, which is consistent with the telescope point spread function measured on standard calibration stars. We measure the nebular continuum to be $\sim 3\%$ of the pulsar continuum in a $2\arcsec \times 0\farcs275$ slice, or about $1.1 \times 10^{-15}$~ergs~cm$^{-2}$~s$^{-1}$~arcsec$ ^{-2}$.
	
At radio wavelengths, the Crab Pulsar has the property of superpulses centered on the main pulse for one random period out of every forty (\cite{lun95}). These superpulses range from 33 to 2000 times the average main pulse. The measured STIS light curve has a raw count rate for the Crab Pulsar of about two events per individual period. Essentially each event is associable with either the peak or the interpulse. We looked at the histogram distribution for the time interval in each individual pulse profile around the main pulse. We found the distribution to be essentially Poissonian in character. Artificially injecting a second event in every fortieth main pulse changed the distribution to be noticably different from a Poissonian histogram. We also looked at the entire light curve and found no significant evidence for large pulses. Based on this simplistic test, we detect no superpulses that are several times larger than the average main pulse.    

The spectrum was dereddened using the Savage \& Mathis (1979) standard galactic interstellar extinction curve and increasing values of $E(B-V)$, until the dust absorption feature at $2200$~\AA\ disappeared. A value of $E(B-V)=0.55\pm 0.05$ results in a smooth continuum (Figure 3b). Fitting the dereddened spectrum to a power-law spectrum, we find $ \alpha _{\nu}= -0.3 \pm 0.2 $. This value compares very favorably to the value of $E(B-V)$=0.51 +0.04 -0.03 obtained for the central nebula from IUE and HUT spectra by Blair et al. (1992). Alternatively adopting $E(B-V)=0.51$ gives $\alpha _{\nu}= -0.1 \pm 0.2 $. Consistent with Percival et al. (1993), we find the standard interstellar extinction curve to apply, contrary to the results of  Benvenuti et al. (1980) who argued for a more peaked $2200$~\AA\ bump. This means that a contribution from supernova ejecta to the extinction curve, as suggested by Benvenuti et al. (1980), is not needed. These values for $ \alpha _{\nu} $ are significantly less than the value measured by \cite{hil97} for the only other pulsar observed in the ultraviolet, PSR B0540-69 in the LMC, which has an  $\alpha _{\nu} = -1.6 \pm 0.4 $.
	
The spectrum of the Crab Pulsar includes several weak absorption lines (Table 2).The measured extinction implies that the estimated neutral hydrogen column density along the line of sight is $2.9 \times 10^{21}$ cm$^{-2}$, from the relationship given between $E(B-V)$ and $N(H~I)$ by de Boer et al. (1987), which appears to hold for many different lines of sight (\cite{shu85}). From the equivalent widths of the absorption lines, a naive interpretation would lead to significant depletion (~$10^{3}$). However, the Crab is at 2 kpc, and multiple interstellar clouds are likely intervening with currently unknown properties. We must defer to other studies at higher dispersion to better understand the weak absorption line measures. From the ratio $N(Mg I)/N(Mg II)$, it is unlikely that the relatively small equivalent widths of these absorption lines are due to the population of higher ionization states for these elements.

We estimate that the centroids of the Mg I, Mg II and Fe II lines are within 2 pixels, or 170 km s$^{-1}$, of the laboratory positions. We find no evidence of blue-shifted components up to $-1900$ km s$^{-1}$ that would be expected if absorptions originate in the fast-moving supernova ejecta  along line of sight (\cite{dav85}; \cite{nas96}). A fast shell (\cite{che77}) is expected to be too highly ionized to be observed in the spectral region $1600$ to $3200$~\AA\ (\cite{lun86}). Also, there are no detected emission line filaments directly in the line of sight to the pulsar. Therefore, it is most probable that the detected absorbing material is in the intervening interstellar clouds or a pre-existing stellar shell (\cite{nas96} and references therein).	

\placetable{tbl-2}

\section{Conclusions}

The STIS UV time-tag mode provides a capability of separating phase dependent spectra with timing resolution as fine as $125~\mu$s. In this application, the Crab Pulsar is demonstrated to have no major wavelength dependence of features within the pulse profile in the $1600 - 3200$~\AA\ spectral region. The $2200$~\AA\ diffuse feature is consistent with the standard interstellar diffuse extinction; the interstellar line absorptions, low in strength, are consistent with low velocity components indicating that the absorption is not coming from the blue-shifted filaments associated with the Crab Nebula. This does not rule out the possibility that some of the line absorption may arise from a low velocity shell of gas due to the wind of the progenitor star, nor that a fast-moving high-ionization shell, including highly ionized magnesium, surrounds the Crab Nebula. Given the 2 kpc distance, the most likely source of the observed features is the intervening interstellar medium. At high galactic latitudes the gas and dust conditions may vary considerably and significantly influence the observed columns densities. Indeed, observations are needed both further into the ultraviolet for the higher excitation absorption lines and at visible wavelengths to determine the intervening cloud structure.

\acknowledgments
We thank the STIS Instrument Development Team and the STIS Servicing Mission Orbital Verification team for their support in obtaining these observations.
\clearpage

\begin{deluxetable}{crrr}
\tablecaption{STIS Observations of the Crab Pulsar \label{tbl-1}}
\tablewidth{0pt}
\tablehead{
\colhead{OBSERVATION} & \colhead{MODE} & \colhead{START TIME} & 
\colhead{EXPOSURE TIME}	}

\startdata
 & &MJD2450667.+ &s\nl
O45701O10 &ACCUM &0.64494 &3 x 600\nl
O45701C0M &TIME-TAG &0.70478 &2400\nl
O45701C2M &TIME-TAG &0.77198 &2400\nl
\enddata	

\end{deluxetable}

\begin{deluxetable}{crr}
\tablecaption{Interstellar Lines Identified in Crab Pulsar 
Spectrum\label{tbl-2}}
\tablewidth{0pt}
\tablehead{
\colhead{Line Identification} & \colhead{Measured Wavelength} & 
\colhead{Equivalent Width}}

\startdata
Fe II 2587 \AA\ &2588.8 \AA\ &0.27+/-0.20 \AA\nl
Fe II 2600 \AA\ &2600.9 \AA\ &0.40+/-0.21 \AA\nl
Mg II 2796 \AA\ &2795.1 \AA\ &0.79+/-0.25 \AA\nl
Mg II 2804 \AA\ &2804.3 \AA\ &0.55+/-0.27 \AA\nl
Mg I  2853 \AA\ &2854.4 \AA\ &0.93+/-0.33 \AA\nl
\enddata	

\end{deluxetable}



\begin{figure}
\epsscale{1.0}
\plotone{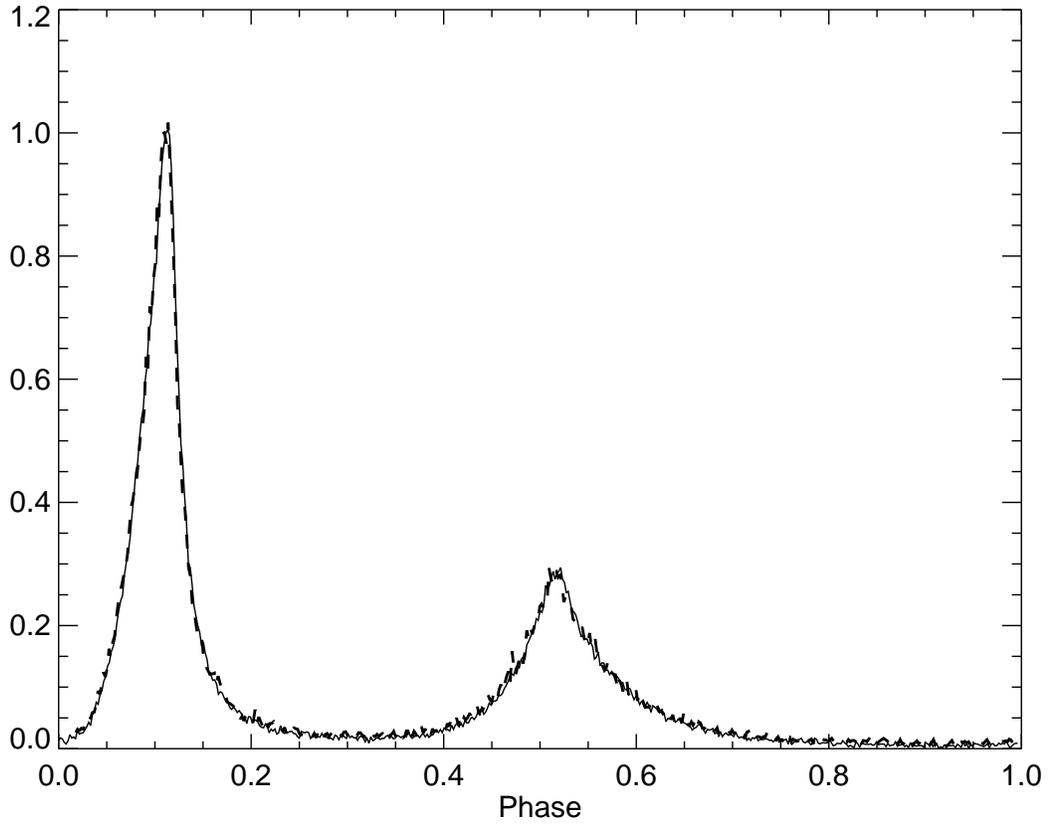}

\caption{Crab Pulsar Pulse Profile: Solid lines indicate the STIS recorded pulse profile. Dashed lines indicate the HSP recorded pulse profile resampled to 512 bins per period (HSP data courtesy of Jeff Percival)\label{fig1}}
\end{figure}

\clearpage

\begin{figure}
\epsscale{1.5}
\plotone{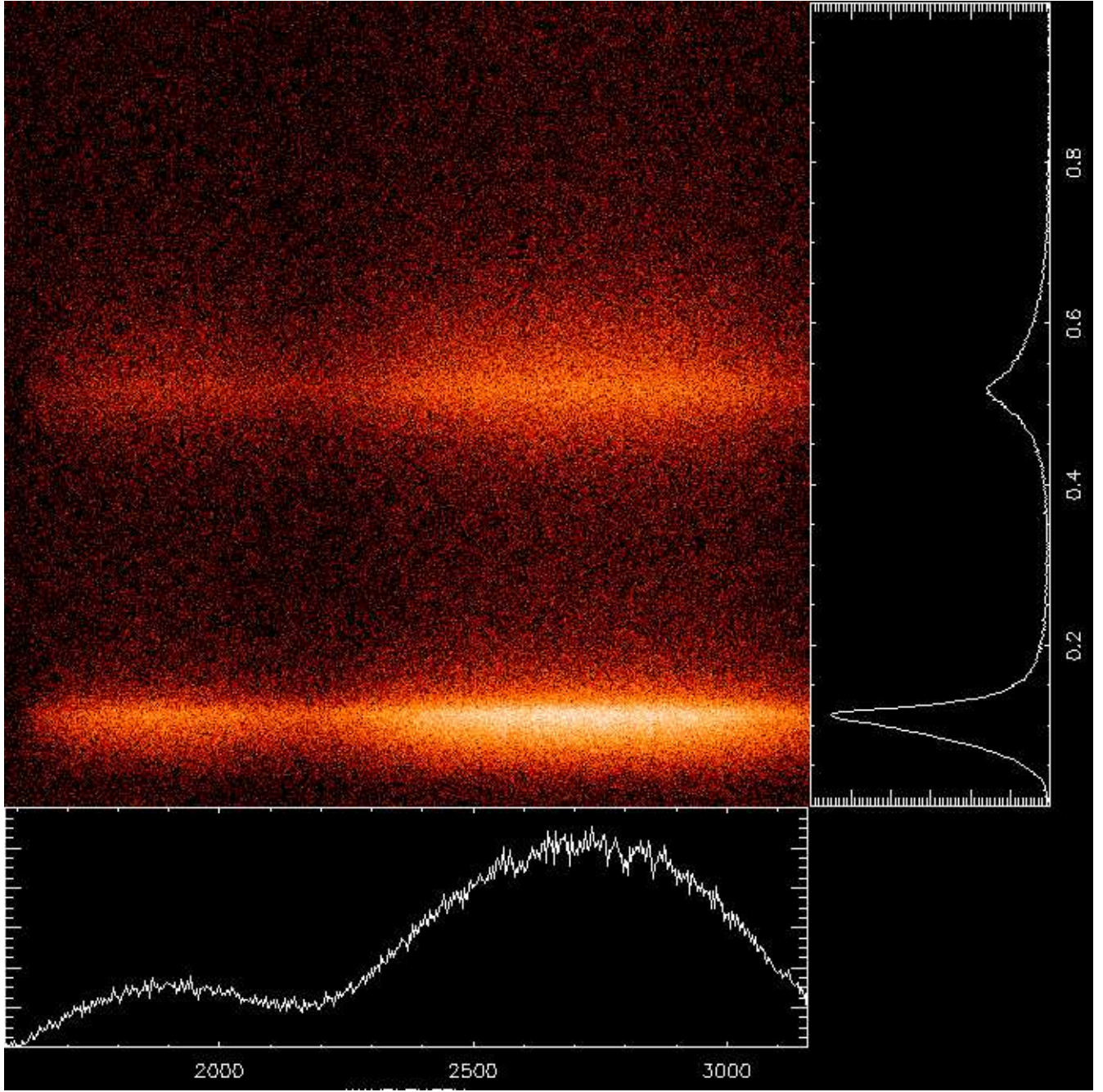}
\caption{Crab Pulsar Pulse Profile versus $1600-3200$~\AA\ Spectral Distribution. Brightness of the two-dimensional plot is proportional to the number of time-tagged events recorded during the two orbit observation. The data do not indicate any wavelength dependence of the pulse profile. 
\label{fig2}}
\end{figure}

\begin{figure}
\epsscale{.9}
\plotone{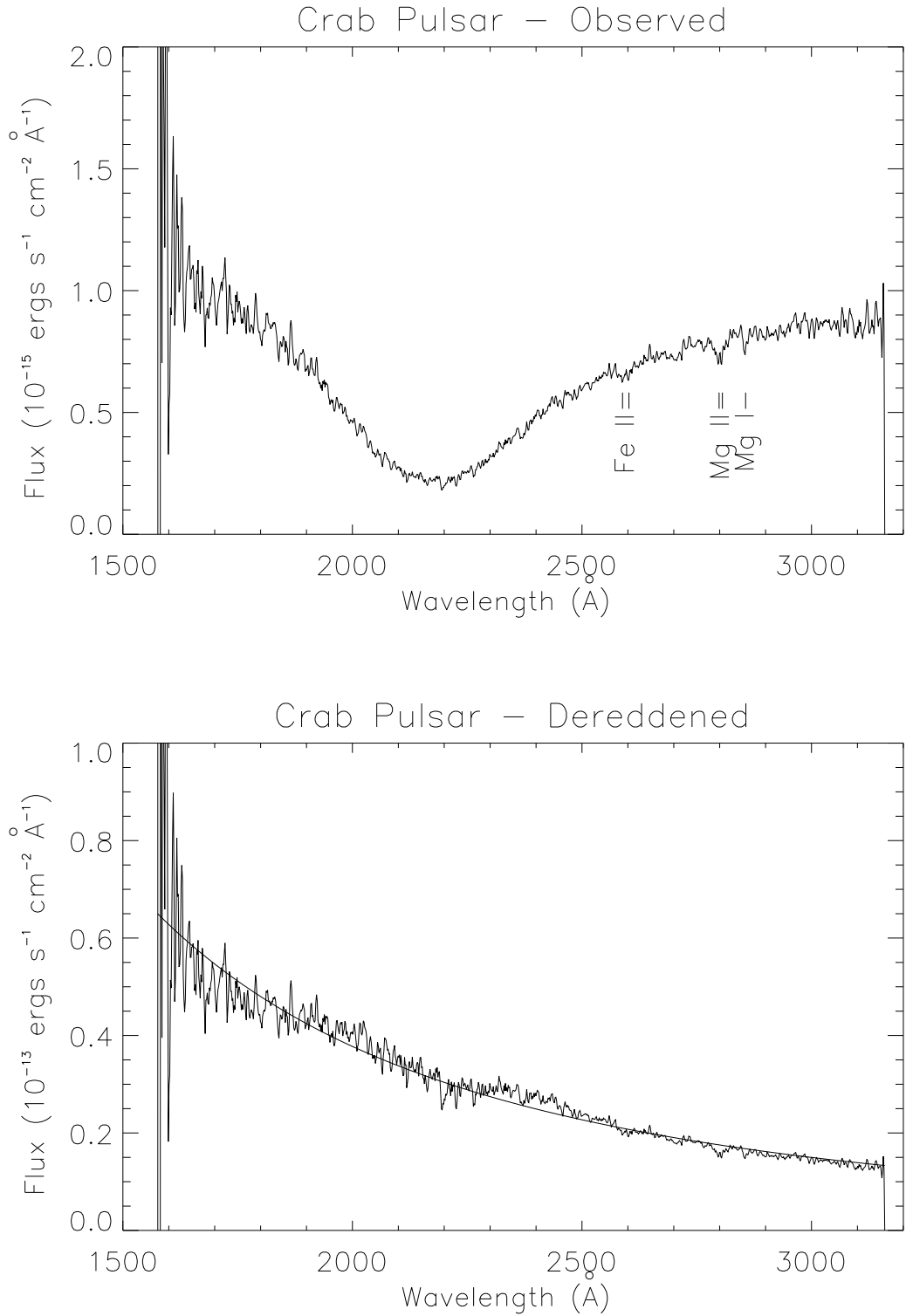}
\caption{Crab Pulsar Spectrum in the range $1600-3200$~\AA. Sum of the entire observations for all three orbits. Figure 3a (top). Flux-corrected spectrum. Figure 3b. Dereddened spectrum using E(B-V)=0.55 and the Savage \& Mathis (1979) galactic extinction curve.
\label{fig3}}
\end{figure}
\end{document}